\documentclass[12pt]{article}
\usepackage{graphicx,floatflt}
\title {THERMODYNAMIC AND SPECTROSCOPIC PROPERTIES  AND\\
 LOW TEMPERATURE THERMOCHROMISM\\ OF CHROMIUM
TRIS--ACETYLACETONATE}

\author{V.N.Naumov, G.I.Frolova, V.V.Nogteva,\\ P.A.Stabnikov,
T.V.Basova, V.A.Nadolinny,\\ Yu.G.Shvedenkov, I.K.Igumenov\\
 Institute of Inorganic Chemistry,\\
Prosp.~Acad.~Lavrentiev 3, 630090 Novosibirsk, Russia}

\begin{document}

\maketitle

\begin{abstract}
Two narrow anomalies with peaks at $\sim 30~K$ and $\sim 60~K$
and a wide diffuse anomaly within the range $110-240~K$ have
been found in a low temperature heat capacity of chromium
tris--acetylacetonate $Cr(AA)_3$. Besides, the reversible change
in color has been discovered when cooling the crystals to liquid
nitrogen temperature (thermochromism).  To clear up the nature
of these effects the static magnetic susceptibility was measured
within the temperature range $2-300~K$, the ESR--spectra on the
$Cr^{3+}$ ion and the transmission spectra in visible region were
recorded at $78~K$ and $300~K$, the Raman spectra were measured
within the range $5-220~K$. It has been ascertained that the
reversible effect of thermochromism is observed in most of
$\beta$--diketonates of the transition metals. Some tentative
considerations concerning the origin of discovered effects are
put forward.
\end{abstract}
\newpage

\section{Introduction}

The chromium tris--acetylacetonate $Cr(C_5H_7O_2)_3$ (or $Cr(AA)_3$)
belongs to the $\beta$--diketo\-na\-te complexes of
transition metals crystallizing in the molecular type lattices.
The $\beta$--diketonates of metals $Me(AA)_3$ are widely used in
practice (to separate metals, to apply metal coatings, as catalysts
etc.). The field of their application is extending with time,
which encourages their further comprehensive investigations.
The compounds $Me(AA)_3$ were actively studied in thermodynamical
[1--3] and crystallochemical \cite{4} aspects. In
the recent years they have became the subject of theoretical
inquiry \cite{5,6}.

When investigating the properties of $\beta$--diketonates we have
found some interesting effects in these compounds, before unknown.
In this work two of such effects have been detected in the
properties of $Cr(AA)_3$: 1) anomalies in low temperature heat
capacity; 2) a reversible change in color of crystals
when cooling them to liquid nitrogen temperature (thermochromism).

\section{Experimental}

The samples $Cr(C_5H_7O_2)_3$ synthesized for this investigation
are crystalline powder with the average size of crystallites
0.3--0.5 mm. At the room temperature they are dark violet. The
samples were defined by the methods of chemical analysis,
IR--spectroscopy, derivatography and X--ray phase analysis.

\begin{figure}[htb]
   \centering
\includegraphics[width=10cm,height=8cm]{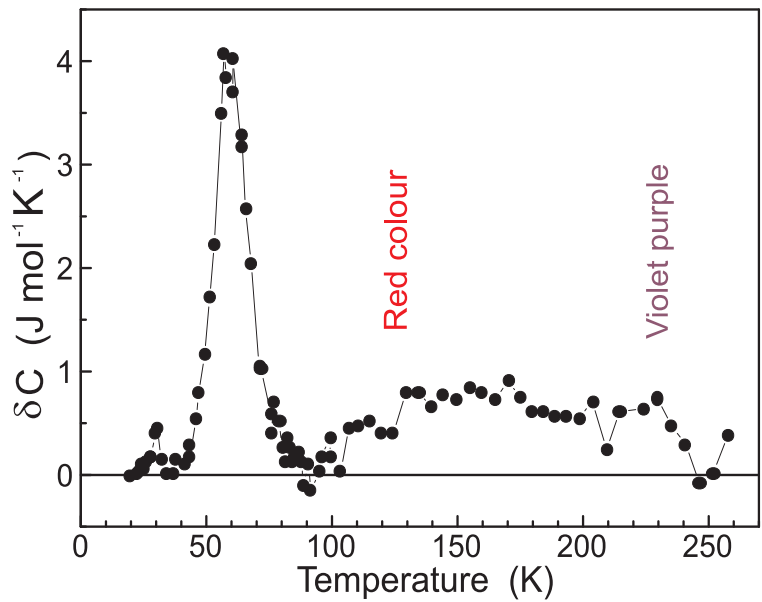}
\caption{\em Anomalous component of heat capacity $\delta C(T)$ of
$Cr(AA)_3$.}
   \label{figdct}
\end{figure}

The heat capacity of $Cr(AA)_3$ has been measured by the adiabatic
method within the range $5-320~K$ using the installation described
in Refs \cite{7,8}. The anomalous component was extracted by
subtracting the regular heat capacity obtained by means of
technique described in \cite{9,10}. Three anomaly were found in
the heat capacity: the small peak at $\sim 30~K$, the anomaly with
maximum at $\sim 60~K$, and the broad anomaly within the range
$110-240~K$ (see Fig.~\ref{figdct}).

The anomalies in maxima are correspondingly 0.9~\%, 3.2~\% and
0.3~\% of a regular heat capacity. The corresponding contributions
to the entropy are: $\Delta S \simeq 0.07 J$ mol$^{-1}K{^-1}$
($30~K$), $\Delta S = 1.20 \pm 0.05 J$~mol$^{-1}K^{-1}$ ($60~K$)
and $\Delta S = 1.5 \pm 0.25 J$~mol$^{-1}K^{-1}$ (the broad
anomaly).

To understand the nature of found anomalies we have measured the
static magnetic susceptibility, the Raman spectra, the ESR--spectra
on the ion $Cr^{3+}$, the transmission spectra in visible region and
have followed the change in color with temperature.

\begin{figure}[htb]
   \centering
\includegraphics[width=10cm,height=7cm]{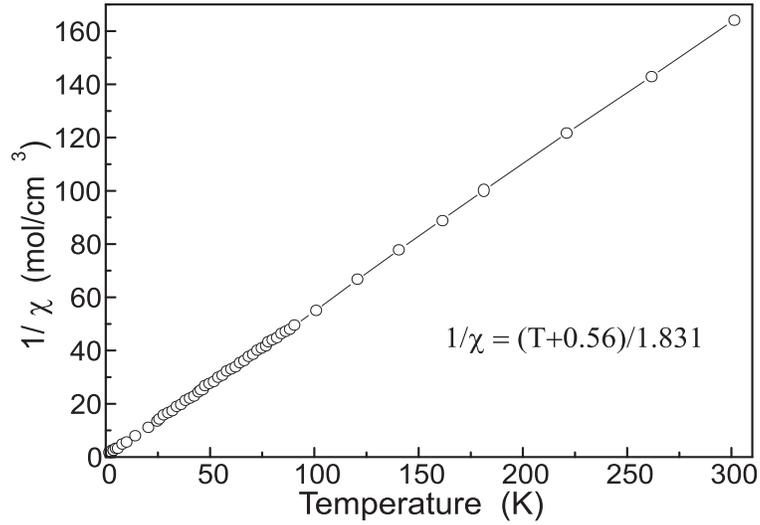}
\caption{\em Magnetic susceptibility $\chi$ as a function of
temperature for $Cr(AA)_3$.}
   \label{fig1xit}
\end{figure}

The static magnetic susceptibility of $Cr(AA)_3$ was measured
by MPMS--5s SQUID--magnitometer (of Quantum Design)
within the temperature range $2-300~K$ (Fig.~\ref{fig1xit}).
The obtained data showed the experimental points
to fit in well with the Curie--Weiss low.
No pronounced magnetic anomalies were observed.
But at the attentive consideration one could see
a weak deviation of experimental points from the Curie--Weiss low
within the temperature range $140-190~K$.

\begin{figure}[htb]
   \centering
\includegraphics[width=10cm,height=7cm]{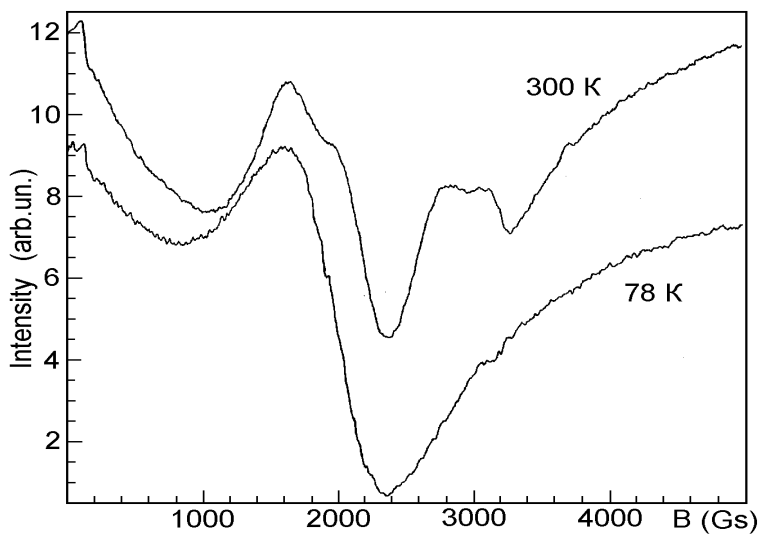}
\caption{\em ESR spectra on $Cr^{3+}$ ion at $78~K$ and $300~K$.}
   \label{figintbgs}
\end{figure}

The ESR--spectra on the ion $Cr^{3+}$
in $Cr(AA)_3$ single crystals
were recorded using $E-109$ Varian spectrometer
at the frequency $9.5$ GHz
and at the temperatures $300~K$ and $78~K$ (Fig.~\ref{figintbgs}).
The ground orbital state of $Cr^{3+}$ ion
in the octahedral surrounding of $Cr(AA)_3$ is singlet;
$g$---factor of $3d^3$ configuration of ion $Cr^{3+}$
is close to $g=2$ \cite{11}.
The large anisotropy of ESR--spectra has been observed
which results from large shifts of spectra
due to the large second order corrections
to the parameters of fine structure
when the initial split is comparable with Zeeman interaction.
The decrease of temperature from $300 K$ to $78~K$
results in the broadening of lines
without any change of their positions, for some lines such a
broadening results in their total disappearance.

\begin{figure}[htb]
   \centering
\includegraphics[width=10cm,height=7cm]{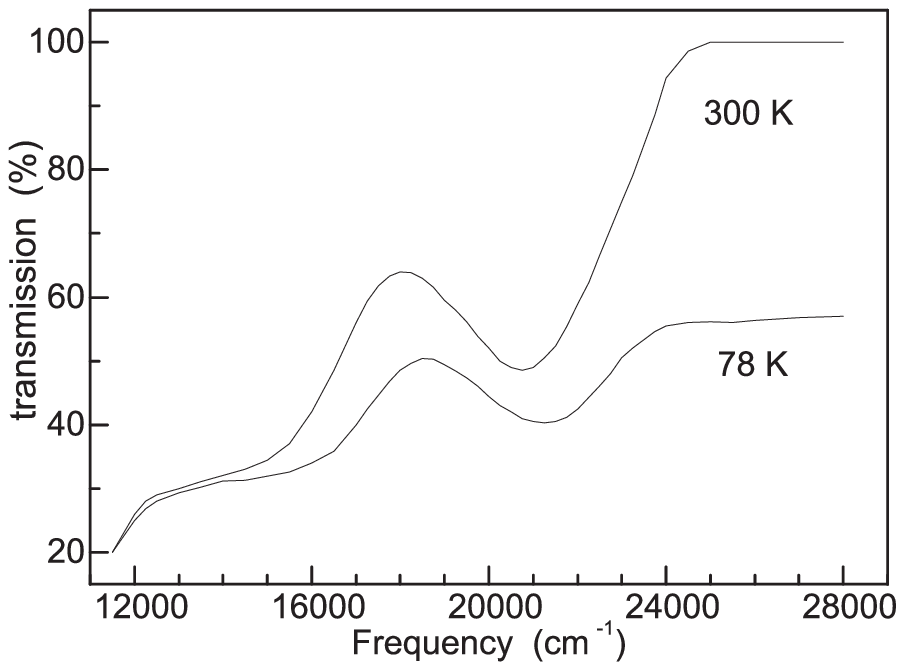}
\caption{\em Transmission spectra of a thin layer of $Cr(AA)_3$ ??in the
visible region?? at 78 K and 300 K.}
   \label{figtrfr}
\end{figure}

The transmission spectra of $Cr(AA)_3$ within the frequency
range $11000-30000$ cm$^{-1}$ have been obtained at room and nitrogen
temperatures using the two--ray spectrophotometer Specord of $UV-200$
type ( Fig.~\ref{figtrfr}). The sample was prepared of powder
placed between two quartz plates and then heated up to
$Cr(AA)_3$ melting temperature ($489~K$). After cooling the
solid continuous layer of $\sim 0.01$ mm thick was formed.

Comparing the spectra obtained in the green region
($18100$~cm$^{-1}$) at room and nitrogen temperatures one can see the
shift of the transmission band maximum to the violet side by
$600$~cm$^{-1}$. This shift exceeds the expected shift resulting from
the temperature expansion of crystal lattice.

\begin{figure}[htb]
   \centering
\includegraphics[width=10cm,height=8cm]{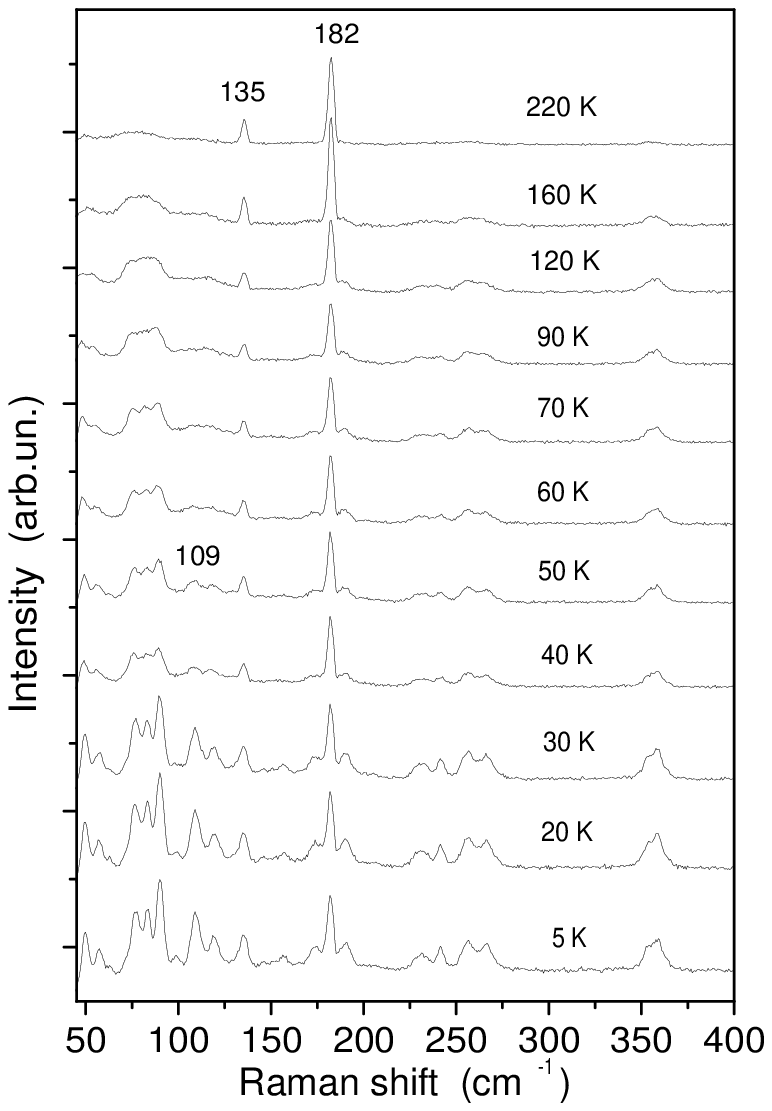}
\caption{\em Raman spectra of $Cr(AA)_3$.}
   \label{figintram}
\end{figure}

The Raman spectra of $Cr(AA)_3$ (Fig.~\ref{figintram}) were
recorded with a Triplemate, SPEX spectrometer equipped with a O--SMA,
$Si$--diode array.  The $633$ nm, $50$ mW line of an $He-Ne$--laser
was used for the spectral excitation. For the low temperature
measurements the sample was fixed on a cold finger of the helium
cryostat (APD Cryogenic Inc). The measurements were carried out
within the temperature range $5-220~K$. The temperature
was established within the accuracy of $0.1~K$.

The frequency interval $40-100$~cm$^{-1}$ is the range of crystal
lattice vibrations. The interval $100-150$~cm$^{-1}$ is the
borderline range of the lattice and molecular vibrations. The
higher frequencies are rated as molecular ones. The band within
the range $40-100$ cm$^{-1}$ being wide at $220~K$ splits up into
the separate components when the temperature decreases. The detail
analysis showed that the frequencies of these components depend on
the temperature, but their relative intensity does'nt depend on
it. When the temperature reaches $\sim 60~K$ the new band at
$109$~cm$^{-1}$ appears which was absent at higher temperature.

It was discovered in this work that the sample
dark--violet at the room temperature becomes red
when immersed into liquid nitrogen (thermochromism). This change in
color is reversible.  In order to follow the change in color with
temperature, a special experiment was carried out. The sample placed
on a copper substrate was immersed in the bath with liquid nitrogen.
Temperature of the substrate and sample was measured by
copper--constantan thermocouple, the color of the sample was
estimated by eye. When the temperature slowly increased, the red
color had remained within the range $78-120~K$. The color of the
sample varied smoothly within the range $120-210~K$, at $210~K$ it
had turned dark violet and then it stayed as such at the further
increase of temperature up to room one.

An additional experiment with 36 compounds showed that the
reversible effects of thermochromism was
observed for most of $\beta$--diketonates of transition
metals, when cooling them with the liquid nitrogen. Most of tried 36
compounds grew light when the temperature decreased. The most
pronounced changes in color were observed for cobalt (III), copper
(II) and chromium (III) compounds. The compound $Co(AA)_3$, being
dark green at room temperature, became violet. The compound
$Cu(AA)(hfa)$--copper(II) acetylacetonate --
hexafluoroacetylacetonate, being green at room temperature, became
blue. The change in color of $Cr(AA)_3$ has been described above.

\section{Discussion}

The nature of discovered anomalies in heat capacity is still
unknown.

For the moment we have no additional information about
the anomaly at $30~ K$.

The $\lambda$--like anomaly at $\sim 60~K$
might display some long range ordering in the substance.
Its temperature is close to temperature $T_s$, where the total
entropy of $Cr(AA)_3$ crystal has the maximal temperature derivative.
The collective long range ordering in solids is often having been
detected at that temperature $T_s$, where the derivative of the total
entropy in temperature is maximal (see, for example, superconducting
effect in HTSC or the ordering of another nature in
Refs.~\cite{12}--\cite{15}).

The magnetic susceptibility data show that the anomalies in the
heat capacity do not result from any magnetic phase transitions.
Then, one can suppose the anomaly to result from the change in the
structure of the crystal or of the molecule. But in Raman spectra
no significant change indicative of structural transformation in
the crystal (of phase transition) was observed. That's why the
crystal is'nt thought to be arranged as a whole. In this case one
may expect a rearrangement or an ordering of the loosely bound
atoms or separate structural fragments: molecules $Cr(AA)_3$,
ligands $C_5H_7O_2$ or groups $CH$ and $CH_3$.

The analysis of the internal vibrations
of the molecule in Raman spectra
has shown no change in structure of molecule with temperature.
The absence of magnetic anomalies testifies that the symmetry of ion
$Cr^{3+}$ crystal surrounding does not change substantially within
the temperature range under investigation.  This means that the
change in geometry of molecule is hardly probable in the vicinity of
chromium ion. One should rather expect some change connected with the
dynamics of the methyl groups.

The torsional vibrations of methyl groups
were identified in Raman spectra
of $Al$--and $Ga$-- $\beta$--diketonates \cite{16}.
They were detected below $77\,K$
in the frequency region $100-140$~cm$^{-1}$.
One may suppose that just as in Ref.~[16]
the new band $109$~cm$^{-1}$
appeared in Raman spectra of $Cr(AA)_3$  below $60~K$ (Fig.5)
results from the scattering
on the torsional vibrations of the methyl groups.
Only these vibrations can show an observed temperature behavior
of the line intensity in Raman spectra.

According to Bose--Einstein distribution the function $\phi(z)$
entering in the general expression for the heat capacity and
characterizing the statistical weight of vibrational modes is
$$
\phi(z)={z^2e^{-z}\over \left(1+e^{-z}\right)^2}\,,\quad
z={\hbar\omega\over kT}.
$$
The maximal change of statistical weight of vibrational mode
$\hbar\omega$ occurs at the temperature $T=\hbar\omega/3k$.
 Then the maximal change of the statistical weight of the
torsion vibrations $109$~cm$^{-1}$ ($\sim 157\,K$) falls at
$\sim 52\,K$.  It is interesting to note that this value is
close to the temperature of the anomaly in heat capacity ($\sim
60~K$).

This $\lambda$--like anomaly (Fig.1) might display some long
range ordering all over the crystal below $\sim 60~K$.  The
torsional vibrations of the methyl groups observed in Raman
spectra at low temperatures arise at the same temperature $\sim
60~K$. The assumed long range ordering might be attributed just
to the dynamics of the methyl groups.

Within the temperature range 110--240 K the diffuse anomaly in
heat capacity takes place.  Within the same temperature range
(120--210 K) the smooth change in the sample color has been
detected.  When temperature decreases from $300~K$ to $78~K$ the
ESR spectra exhibit the broadening of lines without any change of
their energetic position (Fig.3). Such a broadening might be
explained by the appearance of additional indirect anisotropic
interaction which results from the decrease of the mobility of
separate fragments of $Cr(AA)_3$ molecule with temperature. This
broadening is observed within the same temperature range where the
diffuse anomaly in heat capacity and the smooth change in the
sample color take place. The magnetic susceptibility shows the
weak deviation of experimental points from the Curie--Weiss low
within the same temperature range $140-190~K$.

One can suppose that ions $Cr^{3+}$ responsible for the paramagnetic
behavior of $Cr(AA)_3$ crystals might be slightly affected by the
change of some fragments of $Cr(AA)_3$ molecule with temperature. The
change of energetic state of these fragments might result in the
small perturbation of the crystal field. The last results in
redistribution of electron density, shift of the transmission spectra
and change in color of crystals when temperature varies from liquid
nitrogen to room one.

\section{Conclusion}

In the present study some effects have been discovered
in the low temperature properties
of chromium tris--acetylacetonate $Cr(AA)_3$,
which have not been observed before:
the effect of thermochromism
and three anomalies in heat capacity.

Anomalies in heat capacity generally display some transformations in
a substance.

There are no considerations as regards the anomaly at $30~ K$.

As for the $60~K$--anomaly, the assumption has been put forward
that it might display some long range ordering
taken place in the $CH_3$ -- subsystem at this temperature.
The rise of torsional vibrations of $CH_3$ -- methyl groups
at the same temperature was the basis for this assumption.

Besides, it was detected that the wide diffuse anomaly
in the temperature range $110-240 K$ was accompanied with some
features in ESR-- and transmission spectra, with weak deviation
of the magnetic susceptibility from the Curie--Weiss low, and
change in color of the sample in the same temperature range.
All these features were assumed to be attributed to a change
with temperature of some fragments of the $Cr(AA)_3$ molecule
which are still to be recognized.

To verify our tentative insight into the nature of discovered
effects the complex investigation of $\beta$--diketonates of
other transition metals with different ligands should be carried
out.

\section*{Acknowledgements}

The authors thank A.A.Rastorguev for helpful discussions
of essential aspects of this work.

This work is supported by RFBR (grant No 99--03--33370).

\end{document}